\documentclass{aa}
\usepackage{psfig}
\usepackage{graphicx}
\def\vlsr {\hbox{${v_{\rm LSR}}$}}

\def\TEX {$T_{\rm ex}$}
\def\TMB {$T_{\rm MB}$}
\def\TKIN {$T_{\rm kin}$}

\def\TROT {$T_{\rm rot}$}

\def\TASTAR {$T_{\rm A}^{*}$}

\def\H0 {$H_{\rm o}$}

\def\solmass {\hbox{M$_{\odot}$}}
\def\solum {\hbox{L$_{\odot}$}}

\def\numd {\hbox{$n\,({\rm H}_2$)}}                   

\def\kms {\hbox{${\rm km\,s}^{-1}$}}
\def\kmsyr {\hbox{${\rm km\,s}^{-1}\,{\rm yr}^{-1}$}}
\def\kmspc {\hbox{${\rm km\,s}^{-1}\,{\rm pc}^{-1}$}}

\def\Kkms {\hbox{${\rm K\,km\,s}^{-1}$}}
\def\percc {$\hbox{{\rm cm}}^{-3}$}    
\def\cmsq  {$\hbox{{\rm cm}}^{-2}$}    
\def\arcsec {\hbox{$^{\prime\prime}$}}
\def\arcmin {\hbox{$^{\prime}$}}

\def\MOLH {\hbox{${\rm H}_2$}}                    
\def\AMM {\hbox{${\rm NH}_{3}$}}                  
\def\TCO {\hbox{${\rm ^{12}CO}$}}                 
\def\CEIO {\hbox{${\rm C}^{18}{\rm O}$}}          
\def\THCO {\hbox{$^{13}{\rm CO}$}}                
\def\CSEO {\hbox{${\rm C}^{17}{\rm O}$}}          
\def\CTHFOS {\hbox{${\rm C}^{34}{\rm S}$}}        
\def\WAT {\hbox{${\rm H}_2{\rm O}$}}              
\def\METH {\hbox{${\rm CH}_3{\rm OH}$}}           
\def\CH3C2H {\hbox{${\rm CH}_3{\rm C}_2{\rm H}$}} 
\def\HCOP {\hbox{${\rm HCO}^+$}}                  
\def\HTCOP {\hbox{${\rm H^{13}CO}^{+}$}}          
\def\greekg1 {(\zeta _{\rm i},\eta _{\rm j})}

\begin{document}

\title{Molecular Gas and Star Formation in \object{Lynds\,870}
        \thanks {Based on observations with the 10-m Heinrich-Hertz-Telescope
        (HHT) and the MPIfR 100-m telescope at Effelsberg. The HHT is 
        operated by the Submillimeter Telescope Observatory on behalf of 
        Steward Observatory and the Max-Planck-Institut f{\"u}r Radioastronomie.
        The 100-m telescope is operated by the Max-Planck-Institut f{\"u}r
        Radioastronomie.
        }}

\author{R.Q. Mao\inst{1,2,3,4}, J. Yang\inst{1,2},
C. Henkel\inst{3}, Z.B. Jiang\inst{1,2}}

\offprints{R.Q. Mao: rqmao@jets.pmo.ac.cn}

\institute{
  Purple Mountain Observatory, Chinese Academy of Sciences, Nanjing 210008 
\and
  National Astronomical Observatories, Chinese Academy of Sciences, Beijing 100012 
\and
  Max-Planck-Institut f{\"u}r Radioastronomie,
  Auf dem H{\"u}gel 69, D-53121 Bonn
\and
  Institute of Astronomy and Astrophysics, Academia Sinica, P.O. Box 23-141, Taipei 106
}

\titlerunning{Molecular Gas and Star Formation in \object{Lynds\,870}}

\authorrunning{R.Q. Mao, et al.}

\date{Received date / Accepted date}

\abstract{
We present molecular line and submillimeter dust continuum observations of 
the \object{Lynds\,870} cloud in the vicinity of \object{IRAS\,20231+3440}. 
 Two submillimeter cores, SMM1 and SMM2, are identified mapping the 
870\,$\mu$m dust continuum and ammonia (\AMM) emission. The total molecular 
mass is $\sim$70 -- 110\,\solmass. The northern core is warmer and denser 
than the southern one. Molecular outflows are discovered in both cores. 
In the northern one a significant amount of low velocity (1.3--2.8\,\kms) 
outflowing gas is found, that is hidden in the relatively broad CO lines 
but that is revealed by the narrower \HCOP\ spectra. While \object{IRAS\,20231+3440} 
is most likely the exciting star of the northern outflow, the driving source of 
the southern outflow is not detected by infrared surveys and must be deeply 
embedded in the cloud core. Large scale ($\sim$0.2\,pc) infall motion is 
indicated by blue asymmetric profiles observed in the \HCOP\ $J$ = 3--2 
spectra. Red $K_{\rm s}$ band YSO candidates revealed by the 2MASS survey 
indicate ongoing star formation throughout the cloud. The calculated masses 
and the measured degree of turbulence are also reminiscent of clouds forming 
groups of stars. The excitation of the molecular lines, molecular abundances, 
and outflow properties are discussed. It is concluded that \object{IRAS\,20231+3440} 
is a Class\,I object, while the southern core most likely contains a 
Class\,0 source. 
\keywords{
   ISM : jets and outflows -- ISM : molecules -- stars : formation
   ISM : individual (\object{Lynds\,870}, \object{IRAS\,20231+3440})}
}
\maketitle

\section{Introduction} 
\label{introduction}
Stars form from collapsing dense cores of molecular clouds. Protostars 
arise as a result of the collapse and evolve (if their mass does not exceed
a few solar masses) from Class\,0 to Class I, II and III objects (Andr\'e 
et al. \cite{Andre00}). The earliest stages of star formation are observationally 
characterized by an association with dense molecular cores, masers and molecular 
outflows. The dust and the gas of dense cores can be traced by submillimeter continuum 
and spectral line emission from tracers like \AMM, CS, and \HCOP. Molecular outflows, 
usually traced in lines of the CO molecule with broad-line wing emission, have 
received considerable attention over the past two decades and over 250 outflow 
sources have been cataloged so far (Fukui et al. \cite{Fukui93}; Wu et al. \cite{Wu96}). 
Our current understanding of outflow properties and its interaction with the 
parental dense core is, however, heavily influenced by a few well-studied examples, 
and may be far from accurate. 

Recently an IRAS-selected CO $J$ = 1--0 survey has been carried out at 
the 13.7m mm-wave telescope of the Purple Mountain Observatory at Delingha 
(Yang et al. \cite{Yang99}, \cite{Yang02}; Jiang et al. \cite{Jiang00}).  
A number of sources with strong CO emission and/or outflowing gas have been 
selected from the survey. In this work, we present detailed observations of 
one of the sources, \object{IRAS\,20231+3440} (hereafter I\,20231). 

I\,20231 ($\alpha_{\rm 2000.0}$=20$^h$25$^m$07$^s$.0, 
$\delta_{\rm 2000.0}$=34$^{\circ}$50$'$05$''$) is located within \object{Lynds\,870}. 
On Palomar Observatory Sky Survey (POSS) prints, there is no optical counterpart at 
the position of I\,20231. High extinction can be traced over a region of 
15\arcmin$\times$15\arcmin. The infrared colour indices of I\,20231 show that the 
source is deeply embedded (Emerson \cite{Emerson88}). At the same time the IRAS colors
satisfy the criteria introduced by Wood \& Churchwell (\cite{Wood89}) for compact 
HII regions. There were some water maser detections reported (Palla et al. 
\cite{Palla93}; Brand et al. \cite{Brand94}), which indicate an early stage of 
young stellar object (YSO) formation, but methanol (van der Walt et al. \cite{Walt96}) 
and OH masers were not detected (Slysh et al. \cite{Slysh97}). CS 2--1 observations give 
\vlsr\,$\sim$5.4\,\kms (Bronfman et al. \cite{Bronfman96}), which indicates a dynamical 
distance of 3.7 kpc (far) or 1.0 kpc (near distance). The optical extinction seen on 
POSS prints suggests that the nearer distance is more reasonable and we will take 
1.0~kpc as the distance to the source. Since all mapping observations were 
carried out w.r.t I\,20231, we refer in the following to this as the reference position 
(offset (0$''$, 0$''$)).

In this paper, we present molecular line and 870\,$\mu$m continuum observations to 
trace the dense molecular gas, the outflows and the submillimeter dust continuum 
around I\,20231. In \S\,\ref{observations}, we describe our observations, and in 
\S\,\ref{dust} and \S\,\ref{line} we derive 
physical parameters from our dust continuum and molecular line data. 
In \S\,\ref{2mass}, results
from NIR photometry are studied and in \S\,\ref{discussion}, 
we discuss the global properties of star  
formation in the region. Conclusions are drawn in \S\,\ref{conclusion}.

\section{Observations} 
\label{observations}
\subsection{\AMM\ and \WAT\ at Effelsberg 100-m}
\label{eff}
The \AMM\ and the 22\,GHz \WAT\ maser observations were obtained in July 1999, April 
and July 2000, using the Effelsberg 100-m telescope of the Max-Plank Institute
f{\"u}r Radioastronomie (MPIfR) equipped with a maser receiver in 1999 and a dual 
channel K-band HEMT receiver in 2000. The typical system temperature was about 
150\,K -- 200\,K on a main beam brightness temperature scale; the half-power beam 
width was $\sim$40$''$. The data were recorded with a 8192-channel autocorrelator. 
The K-band HEMT receiver allowed us to measure the ($J,K$) = (1,1) to (4,4) \AMM\ 
inversion lines simultaneously. The observed spectral resolution was 0.24 and 
0.98\,\kms\ for \AMM\ observations in 1999 and 2000, respectively, and about 
0.13\,\kms\ for the \WAT\ maser observations. The line intensities were calibrated 
relative to the continuum of \object{NGC\,7027}, for which we assume a flux density 
of 5.86\,Jy, corresponding to a main beam brightness temperature of 8.1\,K. Pointing 
was checked every hour on nearby continuum sources and was found to be stable within 
10$''$. The calibration accuracy is estimated to be $\pm$20\%. 

\subsection{mm and sub-mm line observations at HHT 10m}
\label{HHTline}
Molecular line emission of CO, CS, \HCOP\ and some of their rare isotopes was 
observed with the 10-m Heinrich Hertz Telescope (HHT; see Baars et al. \cite{Baars99}) 
on Mt. Graham in Southern Arizona during three sessions in April 1999, January 2000 
and May 2000. {\it The receivers were sensitive to both sidebands}. A dual channel 
345\,GHz SIS receiver was used for the $J$ = 3--2 transitions of \TCO\ and its isotopes, 
while all the mm-wave lines were observed with a single channel 230\,GHz SIS receiver. 
 For the $^{12}$CO mapping we used the spectral line On--The--Fly (OTF) method, 
in which we scanned the telescope at a rate of about 1$''$ per second. The 
off position was located at $\Delta\alpha=1^{\circ}$ west of I\,20231 and was free 
of CO emission. The most central region ($\sim$1$'$--2$'$) around I\,20231 was also mapped 
in \THCO, \CEIO, \CSEO, \HCOP, \HTCOP and CS with standard position switching (off
position at $\Delta\alpha$ = 20$'$). The $J$ = 5--4 lines of three rare CS 
isotopes, $^{12}$C$^{34}$S, $^{12}$C$^{33}$S and $^{13}$C$^{32}$S, were obtained
towards the central position of I\,20231.  

 For the \TCO\ observations we used two 2048-channel 1\,GHz Acousto-Optical 
Spectrometers (AOS) as backends, with a mean channel spacing of 917\,KHz corresponding 
to $\sim$0.78 and 1.19\,\kms\ at 345 and 230\,GHz, respectively. A 256-channel filter 
bank with a channel spacing of 250\,kHz, providing higher spectral resolution, was 
used simultaneously for other molecular lines. The beam widths were about 22 and 
33\arcsec\ for 345 and 230\,GHz respectively. 

All results displayed are given on a main beam brightness temperature scale (\TMB), 
which is related to \TASTAR\ via \TMB\,\,= \TASTAR\ ($F_{\rm eff}$/$B_{\rm eff}$) 
(see Downes \cite{Downes89}). Main beam efficiencies, $B_{\rm eff}$, were 0.5 and 
0.78 at 345 and 230\,GHz, as obtained from measurements of Saturn.  Forward hemisphere 
efficiencies, $F_{\rm eff}$, were 0.9 and 0.95, respectively. For \TCO, 
\object{$\chi$\,Cyg} was used for both pointing and absolute line intensity calibration. 
The rms pointing uncertainty was found to be $\sim$5\arcsec. Line intensities were found 
to be accurate within 20\% when compared to the $J$ = 3--2 line of CO of Stanek et al. 
(\cite{Stanek95}) and the $J$ = 2--1 line of Loup et al. (\cite{Loup93}). For other lines, 
\object{Orion-KL} was observed as a secondary calibrator and the line survey results by 
Sutton et al. (\cite{Sutton85}) and Blake et al. (\cite{Blake86}) near 230\,GHz and 
Schilke et al. (\cite{Schilke97}) near 345\,GHz were used for comparison. Again calibration
errors should not exceed 20\%.

\subsection{870\,$\mu$m  continuum observations}
\label{HHTdust}
870\,$\mu$m dust continuum emission was imaged with the HHT 19 channel bolometer array. 
The array covers a hexagonal 200$''$ field of view. The beam width of the HHT is 
about 22$''$ at 870\,$\mu$m. Maps were taken using the OTF method with a beam throw of 
120$''$ and a scanning speed of 8$''$ per second. Five OTF maps, with three coverages 
(400$''\times$300$''$ in size) centered at I\,20231 and another two coverages (320$''\times$260$''$) 
centered at (0$''$,--180$''$) relative to I\,20231 were finally combined by mosaicing. The 
average r.m.s noise level is about 90\,mJy in the combined map. All maps were
calibrated using 870\,$\mu$m skydips and the standard calibrators \object{Uranus} and 
\object{G\,34.3+0.2}, and a peak flux density of 60\,Jy\,beam$^{-1}$ was assumed for the latter one.
At 870\,$\mu$m, the typical atmospheric opacity was about 0.4 during the observations.
The calibration accuracy should be better than $\pm$20\%. 

\section{Dust Continuum Emission}
\label{dust}

\begin{figure}
\vspace{0.2cm}
\hspace{0cm}
\psfig{figure=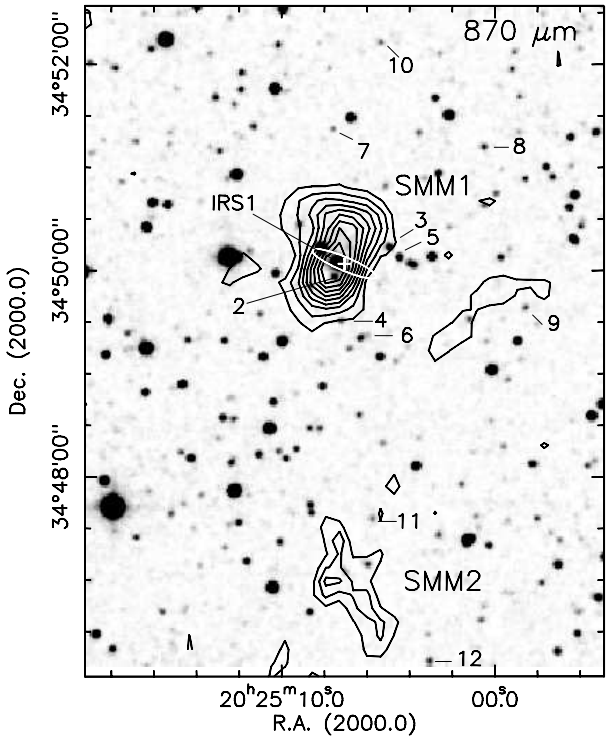,width=8cm,angle=-90}
\caption[]{HHT 870\,$\mu$m continuum image overlaid on a 2MASS-$K_{\rm s}$ image. 
Contour levels are 15\% ($\sim$3$\sigma$) to 95\%  of the peak flux ($\sim$1.7\,Jy)
by steps of 10\%. The position of the IRAS point source I\,20231 is indicated by the 
white cross and the ellipse is the IRAS 1$\sigma$ positional uncertainty. Also shown 
are the 12 red K$_{\rm s}$ sources IRS1, 2 to 12, which are discussed in \S\,\ref{2mass}.}
\label{20231+3440-bolo}
\end{figure}

A 870\,$\mu$m continuum image of L\,870 is presented in Fig.~\ref{20231+3440-bolo}. 
The map reveals two submillimeter cores (SMM), the northern one, SMM1, peaked at 
(7$''$,--7$''$)  with a peak flux density of 1.67\,Jy\,beam$^{\rm -1}$, directly 
associated with I\,20231 and the southern one, SMM2, peaked at (12$''$,--188$''$) with a 
peak flux density of 0.60\,Jy\,beam$^{\rm -1}$, not associated with any IRAS point source.
The northern core region is elongated along a position angle of $P.A.\,\sim$\,--20$^{\circ}$  
with a deconvolved half-power size of 17$''$ by 50$''$ (0.08\,pc by 0.25\,pc), 
while the southern core region looks more irregular. Also shown in the map is a filament 
southwest of I\,20231. This is however only a $\sim$3$\sigma$ detection, and deeper images 
are needed to confirm its existence. The integrated flux densities of the northern, the 
southern and the filament regions are 6.5$\pm$1.1, 1.7$\pm$0.5 and 0.7$\pm$0.2\,Jy, respectively.

The masses of the clouds can be estimated via {\it M=F$_{\rm 
\nu}$D$^{\rm 2}$/$\kappa_{\rm \nu}$B$_{\rm \nu}$(T$_{\rm d}$)}, where $F_{\rm \nu}$ is the 
flux density, $D$ denotes the distance towards the cloud, $B_{\rm \nu}(T_{\rm d})$ is the 
Plank function for temperature $T_{\rm d}$, and $\kappa_{\rm \nu}$ is the dust opacity per 
unit gas and dust mass. Various values of $\kappa_{\rm \nu}$ for different regions can be 
found in the literature (e.g. Draine \& Lee \cite{Draine84}; Hildebrand \cite{Hildebrand83}; 
Preibisch et al. \cite{Preibisch93}; Ossenkopf \& Henning \cite{Ossenkopf94}; Kr\"ugel \& 
Siebenmorgen \cite{Krugel94}). We adopt here the value $\kappa_{\rm 870\mu m}$ = 
0.01\,cm$^{\rm 2}$g$^{\rm -1}$, using the Hildebrand (\cite{Hildebrand83}) assumptions, 
with $\beta$=2 ($\beta$: exponent in the frequency dependence of the dust grain 
emissivity). For the southern dust core, we apply a dust temperature of 12\,K, 
a typical value for  star forming cores (Williams et al. \cite{Williams99}; Bacmann 
et al. \cite{Bacmann00}). For the northern core, we take a higher dust temperature of 
30\,K, as indicated by the color temperature of $\sim$\,33\,K derived from 60 and 
100$\mu$m dust emission (Henning et al. \cite{Henning90}). Total masses traced by the 
dust are then 33.9$\pm$6.3, 40.0$\pm$11.8 and 16.5$\pm$4.7\,\solmass\ for the northern core,
the southern core and the filament, respectively. Note that the masses derived in this way 
can be linearly scaled to another value of $\kappa_{\rm 870\mu m}$ if desired; moreover masses
are sensitive to the temperature and a higher temperature would decrease the calculated mass 
inversely proportional.  

The beam averaged (22$''$) peak column density, derived from the peak flux density of 
1.67\,Jy\,beam$^{-1}$, is 9.6$\times$10$^{22}$\,\cmsq, which gives a corresponding 
peak extinction of $A_{\rm v}$\,$\sim$\,100 ($N(\MOLH)=A_{\rm v}\times 9.5\times10^{20}$\cmsq, 
Bohlin et al. \cite{Bohlin78}). The beam averaged peak column density for a different beam 
size can be calculated by smoothing the image to the corresponding beam size, assuming a 
Gaussian shape of the beam. Column densities of 7.5, 7.0, 6.3 and 5.5$\times10^{22}$\cmsq\ 
are derived for beam sizes of 29, 32, 35 and 40$''$, respectively. To estimate 
molecular abundances assuming they trace the same volume as the 870\,$\mu$m dust continuum 
emission, these values will be used 
in the following.

\begin{table*}
\caption[]{\label{tab:spec-par} Observed molecular line parameters$^{*}$}
\begin{flushleft}
\begin{tabular}{lrccccccc}
\hline
\hline
Line        & Frequency & HPBW     & $T_{\rm MB}$ & \vlsr  & $\Delta$$v_{\rm int}$ & $\int{{T_{\rm MB}}dv}$ & $\bar{\tau}$ & Offset \\
            & (MHz)     & (arcsec) & (K)          & (\kms) & (\kms)      & (K\,\kms)              &              & ($''$,$''$)  \\
\hline
\noalign{\smallskip} 
 $^{12}$C$^{16}$O(3--2) & 345795.975 & 22 &            &             &            & 64.8 (.28) &             & (0,0) \\
 $^{12}$C$^{16}$O(2--1) & 230537.990 & 33 &            &             &            & 82.8 (.32) &             & (0,0) \\
 $^{13}$C$^{16}$O(3--2) & 330587.957 & 23 & 5.37 (.12) &  6.00 (.02) & 3.45 (.04) & 19.8 (.12) & 3.78 (1.54) & (0,0) \\
 $^{13}$C$^{16}$O(2--1) & 220398.686 & 35 & 5.76 (.12) &  6.10 (.02) & 3.68 (.03) & 22.6 (.19) & 3.84 (1.00) & (0,0) \\
 $^{12}$C$^{18}$O(3--2) & 329330.570 & 23 & 2.20 (.15) &  6.17 (.06) & 3.19 (.17) & 8.03 (.30) & 0.50 (0.21) & (0,0) \\
 $^{12}$C$^{18}$O(2--1) & 219560.360 & 35 & 3.31 (.14) &  6.24 (.03) & 2.60 (.06) & 9.26 (.20) & 0.51 (0.13) & (0,0) \\
 $^{12}$C$^{17}$O(2--1) & 224714.368 & 35 & 0.45 (.06) &  6.60 (.15) & 1.65 (.45) & 1.41 (.15) & ...         & (0,0) \\
 $^{12}$C$^{32}$S(5--4) & 244935.643 & 32 & 1.67 (.14) &  6.26 (.05) & 3.23 (.13) & 5.77 (.15) & 1.81 (0.84) & (0,0) \\
                        &            &    & 0.51 (.13) &  6.08 (.10) & 2.72 (.24) & 1.49 (.11) &             & (0,--180) \\
 $^{12}$C$^{34}$S(5--4) & 241016.194 & 32 & 0.13 (.02) &  6.72 (.26) & 2.84 (.55) & 0.39 (.07) & 0.06 (0.03) & (0,0) \\
 $^{12}$C$^{33}$S(5--4) & 242913.680 & 32 & $<$0.06 & & & & & (0,0) \\
 $^{13}$C$^{32}$S(5--4) & 231220.996 & 35 & $<$0.06 & & & & & (0,0) \\
 \HCOP(3--2)     & 267577.625 & 29 &            &             &            & 15.8 (.49) & 3.56 (3.40) & (0,0) \\
 \HTCOP(3--2)    & 260255.480 & 29 & 0.41 (.07) &  6.54 (.08) & 1.85 (.21) & 0.81 (.08) & 0.05 (0.05) & (0,0) \\
                 &            &    &            &             &            &            &             &       \\
 \AMM(1,1)       & 23694.496  & 40 & 3.80 (.10) & 6.17 (.02)  & 1.95 (.05) & 25.2 (.10) & 1.30 (0.10) & (0,0) \\
                 &            &    & 2.50 (.10) & 5.78 (.03)  & 1.35 (.09) & 10.6 (.10) &             & (0,--180) \\
 \AMM(2,2)       & 23722.631  & 40 & 1.30 (.10) & 6.07 (.07)  & 2.44 (.20) & 5.20 (.10) &             & (0,0) \\
                 &            &    & 0.50 (.10) & 5.95 (.13)  & 1.35 (.16) & 0.95 (.30) &             & (0,--180) \\
 \AMM(3,3)       & 23870.130  & 40 & 0.48 (.02) & 6.33 (.07)  & 3.50 (.22) & 1.94 (.07) &             & (0,0) \\
 \AMM(4,4)       & 24139.417  & 40 & 0.08 (.01) & 6.62 (.36)  & 3.77 (.79) & 0.34 (.06) &             & (0,0) \\
\hline
\end{tabular}
\end{flushleft}
$^{*}$ 
Columns 4 to 7 are from Gaussian fitting, except for CO(3--2), (2--1) and \HCOP(3--2) for 
which only total integrated intensities are given due to self-absorption features in the
spectra. Column 8 gives the `averaged' optical depth over the line profile (see text).  
Offset positions relative to I\,20231 are presented in the last column. Note that column 6 gives the intrinsic
full width to half power (FWHP) line width which has been deconvolved with the corresponding 
spectral resolution. Hyperfine 
structure has been taken into account fitting the \CSEO(2--1) and \AMM\ lines. Upper limits 
to the main beam brightness temperatures are 3$\sigma$. The errors (standard deviations) of all these 
parameters are given in parentheses. 
\end{table*}

\section{Molecular Line Emission}
\label{line}

\begin{figure}
\vspace{0.2cm}
\psfig{figure=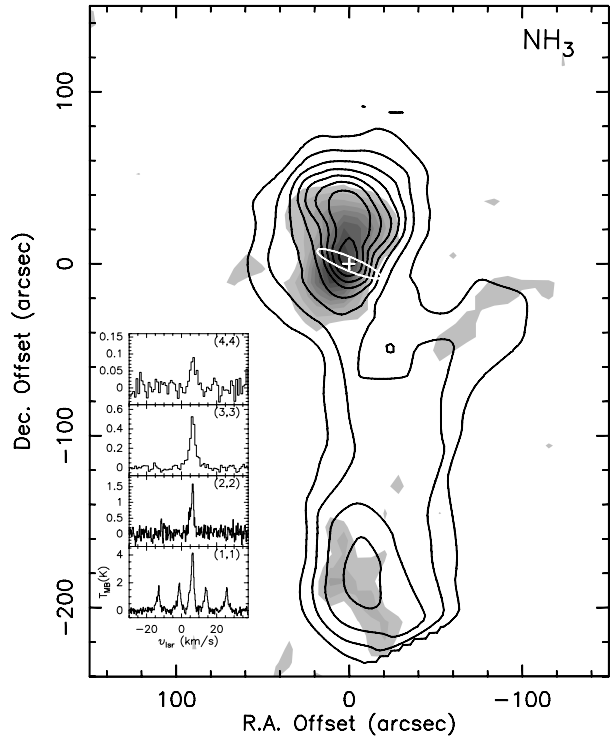,width=8cm,angle=-90}
\caption[]{Effelsberg \AMM\ map (contours) of the integrated (1,1) main group of lines 
overlaid on the 870$\mu$m continuum image (grey). Contour levels are 20\% to 90\% of the 
peak value (9.5\,\Kkms) with a spacing of 10\%. \AMM\,(1,1) to (4,4) spectra taken at the 
I\,20231 position are attached. Note that the spectral resolution is 0.24\,\kms\ for the 
(1,1) and (2,2) lines and 0.98\,\kms\ for the (3,3) and (4,4) lines. For the white cross 
and the error ellipse, see Fig\,\ref{20231+3440-bolo}.}
\label{20231-nh3}
\end{figure}

Results derived from our molecular line observations are summarized in 
Table\,\ref{tab:spec-par}, which lists: molecular line, rest frequency, half power 
beam width (HPBW) at the corresponding frequency, main beam brightness temperature, 
Local Standard of Rest (LSR) velocity, full width to half power (FWHP) line width,
integrated line intensity, estimated optical depth averaged over the line profile, 
and offset position relative to I\,20231. 
The $\Delta$$v_{\rm int}$ values presented in column 6 
are intrinsic line widths that were deconvolved by the spectral resolution.
The $\bar{\tau}$ value given in column 8 is estimated from the ratio of 
integrated intensities of the same transition in two different isotopes,
assuming isotope ratios of [$^{13}$C]:[$^{12}$C]=1:69.7, 
[$^{18}$O]:[$^{17}$O]:[$^{16}$O]=5.5:1:2856 (Wilson \& Rood \cite{Wilson94}), 
and [$^{34}$S]:[$^{32}$S]=31.3 (Chin et al. \cite{Chin96}) for a galactocentric 
distance of 8.2\,kpc. In the case of \AMM, the peak optical depth of the main group 
of hyperfine components is given. We assume local thermodynamic equilibrium (LTE) 
at an excitation temperature \TEX\ and identical beam filling factors (unity) 
throughout the following analysis. For the line emission from linear, rigid 
rotor molecules such as CO, CS, \HCOP\ and their isotopes, the beam averaged 
column density in \cmsq\ is given by
\begin{eqnarray}
\overline{N} = &&10^5\times{{3k^2}\over{4h\pi^3\mu^2\nu^2}}
exp\biggl({{h\nu J} \over 2kT_{\rm ex}}\biggr){{T_{\rm ex}+h\nu/6k(J+1)}
\over{e^{-h\nu/kT_{\rm ex}}}} 
\nonumber \\ 
&&\times{\overline{\tau}\over {1-e^{-\overline{\tau}}}}\int{{T_{\rm MB}}dv}
\end{eqnarray}
(Scoville et al. \cite{Scoville86}), where $\int{{T_{\rm MB}}dv}$ is the integrated
intensity in \Kkms of the $J$+1$\rightarrow$$J$ transitions with frequency $\nu$ (Hz)
and optical depth $\bar{\tau}$. $k$ (erg\,k$^{-1}$)and $h$ (erg\,s) denote the 
Boltzmann constant and the Planck constant, respectively, and $\mu$ (esu\,cm) is 
the permanent dipole moment. For \AMM\ column densities, see \S\,\ref{amm}.

\subsection{\AMM\ data}
\label{amm}

Fig.\ref{20231-nh3} illustrates the distribution of integrated intensity of the 
\AMM\,(1,1) main group of lines. The whole region can be divided into three 
subregions, with two cores located close to the two 870$\mu$m peaks and a ridge 
in between. Such a morphology generally resembles the twin core system defined 
by Jijina et al. (\cite{Jijina99}) in their extensive study of nearby \AMM\ dense 
cores. The northern core has an associated IRAS source (I\,20231), while the southern 
one does not. The spatial structure of the ammonia cores more or less follows the 
870\,$\mu$m dust morphology. With a deconvolved half-power size of 35$''$ by 57$''$ 
(0.17\,pc by 0.28\,pc), the northern core is more centrally peaked than the southern 
one, which has a deconvolved  size of about 75$''$ (0.36\,pc). To the northwest of 
the ridge emission, there is a spur structure, which matches the weak 870\,$\mu$m 
filament, suggesting that the continuum feature is real.

While (1,1) and (2,2) line emission of ammonia is  fairly extended in the region, 
(3,3) emission is  more concentrated around I\,20231 and (4,4) emission is only 
detected towards I\,20231 itself with high signal to noise ratio (S/N). Observed 
parameters of the (1,1) to (4,4) lines are listed in Table\,\ref{tab:spec-par} 
(see attached spectra in Fig.\,\ref{20231-nh3}).

All positions detected in (1,1) and (2,2) emission were analysed with the usual 
technique (see Harju et al. \cite{Harju93}). The rotation temperatures \TROT, the 
column densities $N$(\AMM), and the number densities \numd\ at the two peak positions 
(0$''$,0$''$) and (0$''$,--180$''$) are calculated to be 15.8 and 12.9\,K, 
1.45$\times$10$^{\rm 15}$ and 7.8$\times10^{\rm 14}$\,\cmsq, 1.9$\times$10$^4$ 
and 3.5$\times$10$^3$\,\percc, respectively. However, \TROT\ derived from (1,1) 
and (2,2) emission usually tends to underestimate the gas kinetic temperature 
\TKIN, which can be better approached by a \TROT\ estimated from transitions with
higher excitation (Walmsley \& Ungerechts \cite{Walmsley83}; Danby et al. \cite{Danby88}). 
We thus alternatively apply a rotation diagram for a better determination of \TKIN\ 
for the (0$''$,0$''$) position where a good S/N ratio (4,4) spectrum was taken.
Assuming a single \TROT\ for the populations in all metastable levels, a best fit 
gives a \TROT=31.8$\pm$6.2\,K from the (1,1), (2,2) and (4,4) transitions of 
{\it para-}\AMM. We can thus safely conclude a \TKIN\,$\geq$25\,K for the dense gas 
within a region 40$''$ in diameter around I\,20231.

 From the beam averaged \MOLH\ column density for a 40$''$ area towards I\,20231 (see 
\S\,\ref{dust}), we can estimate the \AMM\ fractional abundance $\chi$(\AMM) = [\AMM]/[H$_2$] 
$\sim$\,2.6$\times$10$^{-8}$. The  masses are then 43, 53 and 12\,\solmass\ 
for the northern core, southern core and the ridge including the spur, respectively. 
These masses agree well with those derived from submillimeter dust emission. The 
total mass of the whole region is therefore $\sim$110\,\solmass. The virial masses 
derived from the line width and core size are 113 and 52\,\solmass\ for the northern 
and southern core, respectively. While the southern core gives a virial mass very 
close to the masses derived from dust and \AMM\ emission, the northern core shows a 
significant inconsistency. This may be due to line broadening by outflowing gas which 
will be discussed below.

\subsection{CO data}
\label{co}

\subsubsection{Ambient CO emission}
\label{ambient}

\begin{figure}
\begin{center}
\vspace{0.2cm}
\hspace{0cm}
\psfig{figure=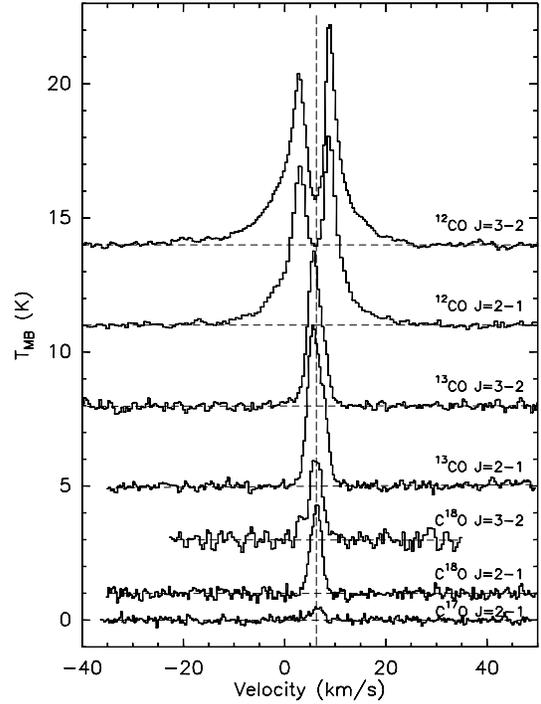,width=7cm,angle=0}
\end{center}
\caption[]{Selected HHT spectra of CO and its isotopes towards I\,20231. The dashed 
vertical line marks the systemic velocity of \vlsr=6.3\,\kms.}
\label{20231-spec-all-tmb}
\end{figure}

Spectra of selected CO isotopes taken at the position of I\,20231 are shown in 
Fig.\,\ref{20231-spec-all-tmb}. The \TCO\ $J$ = 3--2 and 2--1 spectra show similar 
line shapes, with broad-line wings over more than 40\,\kms\ and a deep dip at 
$\sim$6\,\kms. Since the off position was carefully chosen to be free of emission 
and because of the lineshapes of the \THCO, \CEIO\ and \CSEO\ profiles (see below), 
we believe that the dip feature is caused by self-absorption. The self-absorption 
between 3 and 8\,\kms,  with a \TMB\ of about 3\,K and 2\,K in $J$ = 2--1 and 3--2, 
respectively, requires a consistent excitation temperature of \TEX\,$\sim$7.5\,K, 
which is likely the temperature of the outermost layer of the cloud. The \THCO, 
\CEIO\ and \CSEO\ lines are peaked at the velocity of the \TCO\ dip (marked as a 
dashed vertical line in Fig.\,\ref{20231-spec-all-tmb}), with a \vlsr\ of 6.3\,\kms, 
very close to that of our \AMM\ lines. From the ratio of peak temperatures of the 
\THCO\ and \CEIO\ lines, assuming a [\THCO]/[\CEIO] abundance ratio of 7.5 (Wilson 
\& Rood \cite{Wilson94}), we find an LTE excitation temperature of 10\,K for the 
$J$ = 2--1 and 11.5\,K for the $J$ = 3--2 emission, and moderate optical depths of 
0.96 and 0.8 at the peak velocity for the $J$ = 2--1 and 3--2 transitions of \CEIO, 
respectively. We adopt an excitation temperature of 10\,K for the ambient molecular 
gas in the following calculations. Estimated from the ratio of integrated intensities 
of \THCO\ and \CEIO, we find $\bar{\tau}$\,$\sim$0.5 for the $J$ = 3--2 and 2--1 line 
profiles of \CEIO. After applying a $\tau$--correction (multiply by a factor of 
$\bar{\tau}/(1-e^{-\bar{\tau}}$)), we estimate LTE \CEIO\ column densities of 
1.4$\times10^{16}$ and 7.3$\times10^{15}$\cmsq\ for the inner 22$''$ and 33$''$ of 
the cloud. These lead to self-consistent fractional abundances of 
[\CEIO]/[\MOLH]$\sim$1.5$\times$10$^{-7}$ and 1.2$\times10^{-7}$, which are in good 
agreement with the `typical' value of 1.7$\times10^{-7}$ (Frerking et al. 
\cite{Frerking82}). 

\begin{figure}
\begin{center}
\vspace{0.2cm}
\hspace{0cm}
\psfig{figure=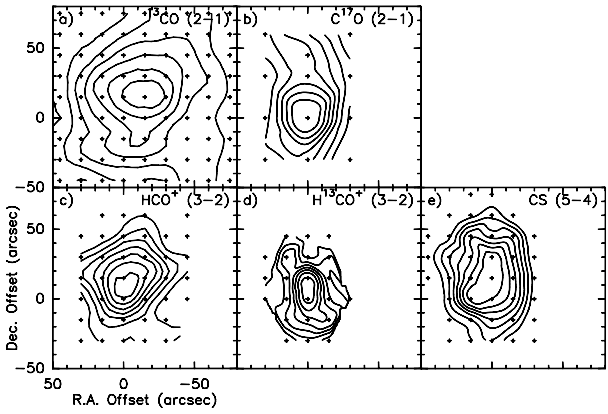,width=8cm,angle=-90}
\end{center}
\caption[]{Contour maps of {\it a}) \THCO\,(2--1), {\it b}) \CSEO\,(2--1), {\it c}) 
\HCOP\,(3--2), {\it d}) \HTCOP(3--2) and {\it e}) CS\,(5--4) taken around I\,20231. 
The contour levels are 
30\% to 90\% by 10\% of the peak values, which are 33.3, 1.4, 17.2, 1.2 and 6.5\,\Kkms, 
respectively. Line names are labeled at the right top corner of each panel. The plus 
signs mark the observed positions.}
\label{20231-allother}
\end{figure}

$^{13}$CO and \CSEO\ were also mapped in the central part of the northern core (see 
panels {\it a)} and {\it b)} of Fig.\ref{20231-allother}). \CSEO\ is believed to be 
optically thin and therefore can be used for mass estimation. The peak LTE column 
density of \CSEO\ is calculated to be 8.6$\times10^{14}$\,\cmsq, which gives a fractional
\CSEO\ abundance of 1.4$\times10^{-8}$. Using the integrated emission over the mapped region, 
we estimate a quiescent molecular cloud mass associated with \CSEO\ emission to be 
$\sim$50\,\solmass, comparable to the mass derived from \AMM. Note that this mass is only a 
lower limit due to the incompleteness of the map. The deconvolved size estimated 
from the \CSEO\ emission is about 40$''$ by 75$''$ (0.2\,pc by 0.36\,pc), slightly larger 
than those estimated from the 870$\mu$m dust emission and the (1,1) line emission of \AMM.

\subsubsection{The outflows}
\label{outflow}

\begin{figure}
\vspace{0.2cm}
\hspace{0cm}
\psfig{figure=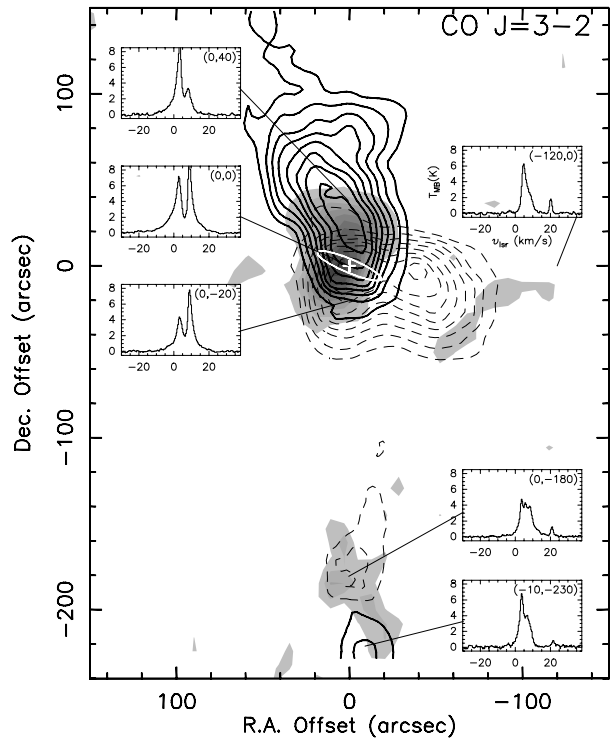,width=8cm,angle=-90}
\caption[]{
Bipolar outflow of CO $J$ = 3--2 (contours) overlaid on our 870$\mu$m continuum image (grey). 
The red-shifted (dashed line) emission is integrated from 9 to 19\,\kms, and the blue-shifted 
(solid line)  emission is integrated from --12 to 3 \,\kms. Contour levels are 30\%, 40\%, 
..., 90\%, 99\% of the peak values, which are  35.2 (blue) and 28.8 \Kkms\ (red), respectively.
Spectra of selected positions (with offsets from I\,20231 in arcsec) are also shown.}
\label{20231-co-bipolar}
\end{figure}

\begin{figure}
\begin{center}
\vspace{0.2cm}
\hspace{0cm}
\psfig{figure=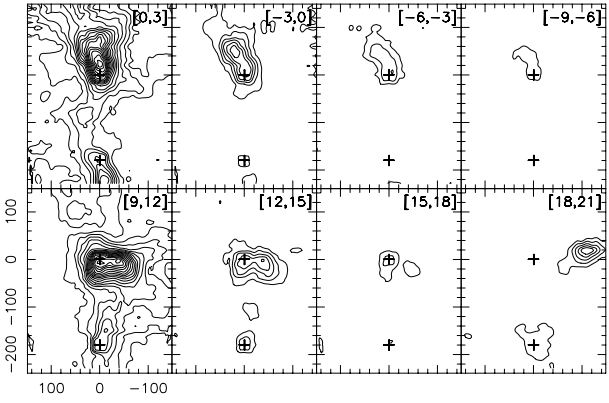,width=8cm,angle=-90}
\end{center}
\caption[]{Velocity channel map of CO $J$ = 3--2. Each plot consists of the intensity integrated 
over a 3.0\,\kms\ range, which is labeled at the right top corner of each plot. The lowest 
contour and the contour spacing are 1.0\,\Kkms. The positions of I\,20231 and (0$''$,--180$''$) 
are labeled as $'$+$'$.}
\label{20231-all-chmap}
\end{figure}

\begin{figure}
\begin{center}
\vspace{0.2cm}
\hspace{0cm}
\psfig{figure=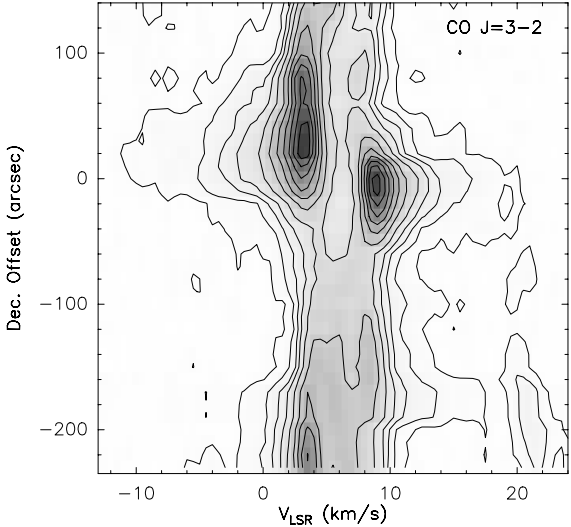,width=7.5cm,angle=0}
\end{center}
\caption[]{Position--velocity map of CO $J$ = 3--2 emission along Dec. for an R.A. offset = 0. 
The contours and intervals are 0.3 to 2.1 by 0.6\,K, and 3 to 9 by 1\,K.}
\label{20231-co32-pv}
\end{figure}

Broad-line wing emission has been detected in both CO $J$ = 2--1 and 3--2 
transitions toward the central region in L\,870. The \TCO\ wing emission, 
determined from a comparison of \CEIO\ and \TCO\ profiles, is defined 
by velocities redder than 9 \kms\ or bluer than 3 \kms. Since CO $J$=2--1 and 3--2 
maps of the outflowing material show very similar bipolar structures, we 
present in Fig.\,\ref{20231-co-bipolar} only the higher angular resolution CO 
$J$=3--2 data. The blue- (--12 to 3\,\kms) and the red-shifted (9 to 19\,\kms) 
lobes are shown in solid and dashed contours, respectively. Note that the 
integral of the redshifted wing emission was intentionally truncated to 
avoid contamination from a separate velocity component at about 20\,\kms, 
which appears in three attached spectra at the right side of 
Fig.\,\ref{20231-co-bipolar}.

Being centered on I\,20231, the outflow lobes are asymmetric and show 
some evidence of clumpiness, especially in the red lobe. There is some overlap 
between red and blue-shifted emission. Interestingly, close to the southern 
dust continuum peak, a pair of weak blue- and red-shifted lobe is also detected, 
indicative of  ongoing star formation activity in this region too. 
Velocity channel maps (Fig.\ref{20231-all-chmap}) and a position velocity 
map along the declination axis (Fig.\ref{20231-co32-pv}) support this view. 
The blue-shifted lobe is, however, not fully covered by our map.

Fig.\,\ref{20231-all-chmap} presents a series of velocity maps in steps of 
3\,\kms. We have skipped the emission between 3\,\kms\ and 9\,\kms\ that is 
affected by self-absorption. As the maps show, the blue emission in the northern
region forms a well-collimated lobe with I\,20231 near one end and extends 
northeast, while the red emission shows two peaks, one at I\,20231 and the 
other 40$''$ to the west. There are two clumps emerging in the 18 to 21\,\kms\ 
channel map (right bottom panel in Fig.\,\ref{20231-all-chmap}), with a more 
compact one located about 2$'$ west of I\,20231 and another one centered around 
the southern dust continuum peak SMM2. They are most likely cloudlets along the 
light-of-sight towards L\,870, but the kinematic distances could not be estimated 
since their characteristic velocities fall into a forbidden zone. As shown in 
Fig.\ref{20231-co32-pv}, the strong self-absorption at ambient velocities in the 
northern core appears between two bright emission peaks at about 3 and 8.5\,\kms. 

The mean optical depth in the line wings, calculated from \TCO\ to \THCO\ line 
ratios, is $\sim$3.0 for blue- (--12 to +3\,\kms) and $\sim$2.0 for redshifted
(9 to 19\,kms) wing emission. From  the ratio of CO $J$ = 3--2 and  2--1 brightness 
temperatures (the 3--2 spectra were smoothed to the angular resolution of the 2--1 
profiles), we estimate the LTE excitation temperature $T_{\rm ex}$ of the outflowing 
gas to be 20\,K to 40\,K for an extended source and 10\,K to 14\,K for a point source. 
We therefore take $T_{\rm ex}$=20\,K for our outflow parameter calculations described 
below.

\begin{table*}
\begin{center}
\caption[]{\label{outflow-par} Physical parameters of the two CO outflows$^{a}$}
\begin{tabular}{l|l|ccccccccc}
\hline
\hline
\multicolumn{2}{c|}{Characteristic} & $R_{\rm flow}$ & $<V>$ & $t_{\rm d}$ & 
$M_{\rm flow} $ & $\dot{M}^{b}$ & $P$ & $F$ & $E$ & $L_{\rm mech}$  \\
\multicolumn{2}{c|}{} & (pc) & ($\kms$) & (10$^4$yr) & ($\solmass$)& 
($\solmass$yr$^{-1}$) & ($\solmass\kms$) & (\solmass\kmsyr) & (10$^{45}$ergs) &  ($\solum$) \\
\hline
\hline
      & Blue  & 0.6       & 10.0       & 6.0       &  1.3 & 2.2$\times10^{-5}$   & 12.7 & 2.1$\times10^{-4}$ & 1.3 & 0.24 \\
North & Red   & 0.4       & 7.8        & 5.1       &  0.8 & 1.6$\times10^{-5}$   & 6.0  & 1.2$\times10^{-4}$ & 0.5 & 0.10 \\
      & Total & 0.5$^{c}$ & 8.9$^{c}$  & 5.6$^{c}$ &  2.1 & 3.8$\times10^{-5}$   & 18.7 & 3.3$\times10^{-4}$ & 1.8 & 0.34 \\
\hline
      & Blue  & 0.3$^{d}$&  7.2       & 4.0         & 0.07  & 1.8$\times10^{-6}$ & 0.50 & 1.3$\times10^{-5}$ & 0.04 & 0.007 \\ 
South & Red   & 0.3      &  7.2       & 4.0         & 0.08  & 2.0$\times10^{-6}$ & 0.58 & 1.4$\times10^{-5}$ & 0.04 & 0.009\\
      & Total & 0.3$^{c}$&  7.2$^{c}$ & 4.0$^{c}$   & 0.15  & 3.8$\times10^{-6}$ & 1.08 & 2.7$\times10^{-5}$ & 0.08 & 0.016\\
\hline
\end{tabular}
\end{center}
${}^{a}$ An inclination angle of 45$^{\circ}$ has been taken into account
for all parameters listed in this table. 
{}$^{b}$ Note the $\dot{M}$ here is the entrainment rate.
{}$^{c}$ The average of two lobes.
{}$^{d}$ Due to lack of coverage in the southern outflow, R$_{\rm flow}$ for 
the blueshifted lobe is assumed to be equal to that of its redshifted counterpart.
\end{table*}

Assuming that the CO emission is near LTE and that the \TCO\ fractional abundance is 
[\TCO]/[\MOLH]$\sim$10$^{-4}$, the outflow parameters were estimated following a method 
discussed by Cabrit \& Bertout (\cite{Cabrit86}, \cite{Cabrit90}). The mass of the outflowing 
gas is estimated from the integrated CO $J$ = 3--2 intensities of the red- and blueshifted 
lobes. The momentum $P$ and the energy $E$ are given by $M_{\rm flow}<V>$ and $M_{\rm flow}<V>^{\rm 2}$/2,
respectively, where $<V>$ is the intensity-weighted velocity. We can also get 
the entrainment mass rate $\dot{M}=M_{\rm flow}/t_{\rm d}$, the force in 
the flow $F=P/t_{\rm d}$, and the mechanical luminosity $L_{\rm mech}=E/t_{\rm d}$, by taking the 
characteristic flow time scale, $t_{\rm d}=R_{\rm flow}/<V>$, where the flow radius,  
$R_{\rm flow}$  is the lobe extension. 
Correcting the flow parameters for an inclination angle (to the line of sight) 
of 45$^{\circ}$, we summarize 
the outflow parameters in Table\,\ref{outflow-par}. The corresponding parameters of 
the southern outflow are estimated assuming optically thin wing emission, since \THCO\ 
data are not available in this case.
 
The parameters given in Table 2 are typical for low-mass or intermediate-mass YSOs. 
Cabrit \& Bertout (\cite{Cabrit90}) estimated typical errors that should approximately 
be within a factor of $\sim$ 3 for $M_{\rm flow}$, a factor of $\sim$ 10 for $F$, and a 
factor of $\sim$ 30 for $L_{\rm mech}$. These errors may be introduced by the uncertainties 
in [\TCO]/[\MOLH], the distance, line excitation temperature, inclination of the flow axis 
to the line of sight, and optical depth of the CO line. By using intensity weighted 
velocities rather than the maximum velocities, the values of $t_{\rm d}$ given in 
Table\,\ref{outflow-par} may be overestimated. The flow radii $R_{\rm flow}$ were 
corrected for an inclination of 45\degr\ which is likely an overestimate for the northern 
outflow, but an underestimate for the southern outflow when we compare our observed line 
profiles, outflow structure and position-velocity maps to those of Cabrit \&\ Bertout 
(\cite{Cabrit90}). In the case of the northern outflow, effects caused by taking `average'
velocities and a `standard' inclination may cancel each other with respect to $t_{\rm d}$.
In the case of the southern outflow, however, they add up so that the $t_{\rm d}$ value
is an upper limit for the southern outflow. We can therefore safely conclude that 
the dynamical time of the northern flow is longer than that of the southern flow.
 
\subsection{\HCOP\ and \HTCOP}
\label{hco+}

\begin{figure}
\begin{center}
\vspace{0.0cm}
\hspace{0cm}
\psfig{figure=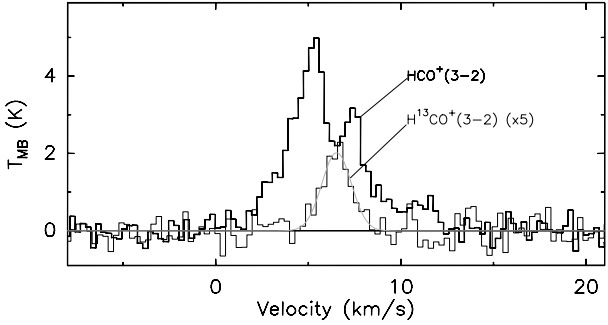,width=8cm,angle=-90}
\end{center}
\caption[]{$J$ = 3--2 spectra of \HCOP\ (dark line) and \HTCOP ($\times$5) (light line) 
taken at I\,20231. The channel spacing is 0.28\,\kms.}
\label{spec-hco+}
\end{figure}

The \HCOP\ and \HTCOP\ $J$ = 3--2 emission lines were measured in the central area of 
the northern core (see panels {\it c)} and {\it d)} in Fig.\,\ref{20231-allother}).
Fig.\,\ref{spec-hco+} shows the spectra taken towards I\,20231. Strong \HCOP\ 
self-absorption is observed at the peak velocity of the \HTCOP\ line ($\sim$\,6.5\,\kms). 
Weak blue- and red-shifted line wing emission is seen in the $J$ = 3--2 line of 
\HCOP\ that can be interpreted as additional evidence for outflowing gas. The asymmetry 
in the line showing enhanced blue-shifted emission characterizes all spectra within a 
radius of $\sim$40$''$ ($\sim$0.2\,pc) around I\,20231. This is indicative of possible
infall motion that will be discussed in \S\,\ref{infall}.

The total integrated intensity map of \HTCOP\ shows a very compact core with a deconvolved 
half-power size of about 20$''$ by 35$''$ (0.1\,pc by 0.17\,pc; see panel {\it d} in 
Fig.\ref{20231-allother}). With the same method used above, we estimate a fractional
\HCOP\ abundance of 9.6$\times10^{-10}$, about an order of magnitude lower than the typical 
value of $\sim$8$\times$ 10$^{-9}$ in dark clouds (Irvine et al. \cite{Irvine87}), but close 
to the inferred value of 1$\times10^{-9}$ for envelopes of submillimeter continuum sources
in the Serpens molecular cloud (Hogerheijde at al. \cite{Hogerheijde99}). This may indicate 
a depletion of \HCOP\ in this region, i.e. some molecules 
are frozen out onto the dust. The mass associated with \HTCOP\ is $\sim$50\,\solmass, 
consistent with the mass derived from \CSEO.

\begin{figure}
\begin{center}
\vspace{0cm}
\hspace{0cm}
\psfig{figure=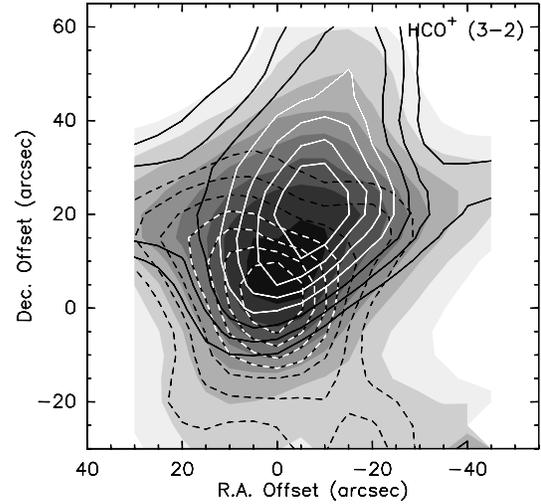,width=7cm,angle=-90}
\end{center}
\caption[]{Bipolar outflow (contours) overlaid on the total integrated intensity map 
(grey) of \HCOP\ $J$ = 3--2 taken around I\,20231. 
The blue- (solid) and the redshifted (dashed) emission
is integrated from 3 to 5\,\kms\ and 7.5 to 9\,\kms, respectively. Contour levels are 
30\%, 40\%,..., 90\% of the peak values, which are 5.7 and 7.2\,\Kkms, respectively.
Grey scales are 4.4 to 19.8 by steps of 2.2\,\Kkms.}
\label{20231-hco+-bipolar}
\end{figure}

As already mentioned, broad-line wings were detected in the \HCOP\ $J$ = 3--2 spectra. The outflowing emission  
shown in Fig.\ref{20231-hco+-bipolar} was integrated over a velocity range of 3 to 
5\,\kms\ for blueshifted (solid contours) emission and 7.5 to 9\,\kms\ for redshifted 
(dashed contours) emission, respectively. By assuming LTE at an excitation temperature 
\TEX\,=20\,K and optically thin outflow emission, we estimate the total mass of outflowing 
gas traced by \HCOP\ in the velocity range mentioned above to be 6\,\solmass, almost three 
times the gas flow traced by CO. This is mostly due to the inclusion of the low velocity 
outflowing gas which could not directly be identified in the broader line profiles of CO. 
We thus find a total mass of 8\,\solmass\ for the outflow. Although such a mass estimate 
depends on uncertain molecular abundances, we can conclude that a significant amount of
the outflowing gas is observed at low velocities and that the total outflowing gas mass
is severely underestimated by the CO data alone.

\subsection{CS and \CTHFOS}
\label{cs}

\begin{figure}
\begin{center}
\vspace{0cm}
\hspace{0cm}
\psfig{figure=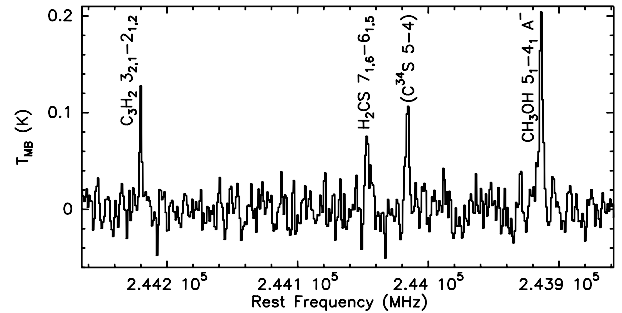,width=8cm,angle=-90}
\end{center}
\caption[]{Dual sideband spectrum with the \CTHFOS\ $J$ = 5--4 line from the signal sideband and
the other lines from the image sideband. Identified molecular transitions are marked.
The channel spacing is 0.62\,\kms.}
\label{spec-iband}
\end{figure}
 
In contrast to CO and \HCOP\, our CS profiles show no sign of self-absorption.
With a critical density of about 7$\times$10$^6$\,\percc, the $J$ = 5--4 transition 
of CS and its isotopes traces the very dense molecular gas. While CS and \CTHFOS\ were 
detected towards I\,20231, C$^{33}$S and $^{13}$CS remain undetected with an r.m.s level of 
$\sim$20\,mK and a channel separation of 0.3\,\kms. A Gaussian fit to the CS $J$ = 5--4
profile at the (0$''$,0$''$) position gives a velocity of 6.3\,\kms, the same as that for
the \CEIO\ $J$ = 2--1 line but different from that of CS\,$J$ = 2--1 (Bronfman et al. 
\cite{Bronfman96}). We believe that this is mainly due to different optical depths, 
i.e., optically thick emission in the $J$ = 2--1 line and optically thin emission in the 
5--4 line. A nine point CS map was also taken towards the southern core with the peak 
located at (0$''$,--180$''$). As shown in Table\,\ref{tab:spec-par}, the CS emission 
lines are weaker and narrower towards the southern peak, and the emission appears to be 
much more compact than that in \AMM. The integrated CS intensity map (Fig.\ref{20231-allother}, 
panel {\it e}) shows a slightly elongated structure in N-S direction with a peak about 
10$''$ north of I\,20231. 

Assuming the same excitation temperature for the $J$ = 5--4 lines of CS and C$^{34}$S, 
we can estimate the CS optical depth for a given isotopic abundance ratio $R$ (e.g. 
Zinchenko et al. \cite{Zinchenko94}). With $R$ = 31.3 for I\,20231 (Chin et al. 
\cite{Chin96}), the excitation temperature, the peak optical depth and the mean optical depth 
averaged over the line profile are estimated to be 6.5\,K, 2.2 and 1.8. We then derive 
a CS fractional abundance of 7.8$\times$10$^{-9}$, and a core mass of about 26\,\solmass. 

While C$^{34}$S $J$ = 5--4 was detected in the lower sideband, \METH\,5(1)--4(1)A$^{-}$ (243.9158\,GHz), 
H$_{\rm 2}$CS\,7$_{\rm 1,6}$--6$_{\rm 1,5}$ (244.0478GHz) and 
C$_{\rm 3}$H$_{\rm 2}$\,3$_{\rm 2,1}$--2$_{\rm 1,2}$ (244.2221GHz) were detected in the 
upper sideband with peak intensities of 0.19, 0.10 and 0.14\,K and integrated intensities 
of 0.63, 0.57, and 0.27\,\Kkms, respectively (see Fig.\,\ref{spec-iband}). Since these 
lines are all from non-linear molecules, equation (1) cannot be used to estimate the column 
densities. Instead, we have used a relationship between the energy of the upper level and 
the integrated line intensity (see e.g. Groesbeck et al. \cite{Groesbeck94}) for an optically 
thin transition, adopting the line strength and partition function information from the JPL 
line catalog (Pickett et al. \cite{Pickett98} and online at {\em http://spec.jpl.nasa.gov}).
Under the assumption of LTE and optically thin line emission, taking \TEX=10\,K, we estimate
CH$_3$OH, H$_2$CS and C$_3$H$_2$ column densities to be 6.6$\times10^{\rm 14}$, 
9.0$\times10^{\rm 13}$ and 1.1$\times10^{\rm 13}$\,\cmsq, and thus fractional abundances to be
9.7$\times10^{-9}$, 1.3$\times10^{-9}$ and 1.6$\times10^{-10}$, respectively. These abundances 
are lower limits. If the lines are not optically thin, column densities and abundances
would be higher. Note that abundances are estimated assuming that molecular lines are 
sampling the same volume as the 870\,$\mu$m dust continuum emission (see \S\,\ref{dust}). This does not 
necessarily hold, since different molecular species and even transitions of the 
same species have fairly different critical densities. Multi-line observations and radiative 
transfer modeling are needed to test this hypothesis for I\,20231.

\subsection{Water Maser Emission}
\label{water}

\begin{figure}
\begin{center}
\vspace{0.2cm}
\hspace{0cm}
\psfig{figure=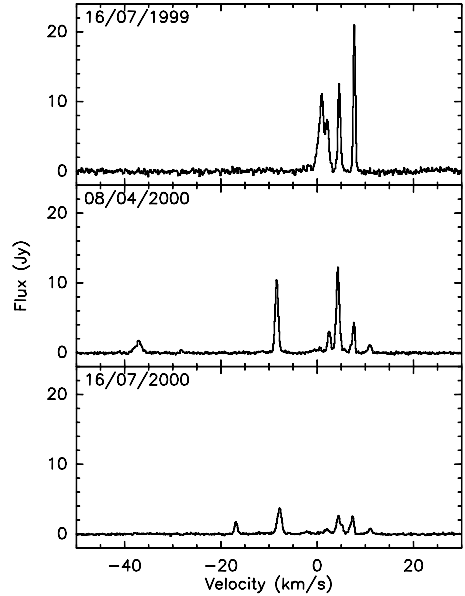,width=6cm,angle=0}
\end{center}
\caption[]{Three 22\,GHz H$_{\rm 2}$O maser spectra taken at I\,20231 with the epoch labeled at the left top 
corner of each panel. The channel spacing is 0.13\,\kms. The full observed velocity range is $\sim$\,--110
 to 140\,\kms. Only those parts of the spectra are shown that contain detected emission.}
\label{20231-h2o-3epochs}
\end{figure}

Fig.\,\ref{20231-h2o-3epochs} shows 22\,GHz water maser spectra taken with the Effelsberg 100-m 
telescope. As shown in the spectra, the velocity ranges of maser emission are between 
--40\,\kms\ and +12\,\kms, with more blue features than red ones relative to the systemic 
velocity of 6.3\,\kms. The maser spectra also show time variability on a timescale 
$\leq$3 months. While the total \WAT\ maser luminosity decreased by a factor of $\sim$3 
from 40.8\,Jy\,\kms\ in July 1999 to 14.3\,Jy\,\kms\ in July 2000, some individual maser 
components showed even stronger variations. Combining our data with the water maser 
spectra reported by Brand et al. (\cite{Brand94}), the total \WAT\ luminosity 
$L_{\rm H_{\rm 2}O}$, assuming isotropic emission, varied between 3.3$\times$10$^{-7}$ 
and 1.1$\times$10$^{-6}$\,\solum. The upper limit is consistent with what one expects 
for a maser associated with an IRAS source with $L_{\rm FIR}$\,$\sim$10$^{2}$\solum\ 
(Henning et al. \cite{Henning92}). From the mapped maser emission in April 2000, almost 
all the maser components, e.g. those at --37.1, --8.26, 4.33, 7.59 and 11.00\,\kms, are 
centered at the IRAS position within a pointing accuracy of $\sim$5$''$, except the
2.48\,\kms\ feature which peaked at (0$''$,10$''$). We also observed the (0$''$,--180$''$) 
position (i.e. the southern \AMM\ and sub-mm continuum peak), but no signal was seen at an 
rms noise level of 0.1\,Jy (channnel spacing: 0.13\,\kms), which yields an upper limit 
of $L_{\rm H_{2}O}$ $\sim$ 3$\times10^{-10}\,\solum$. 

\section{2MASS NIR photometry}
\label{2mass}

\begin{figure}
\begin{center}
\vspace{0.3cm}
\hspace{0cm}
\psfig{figure=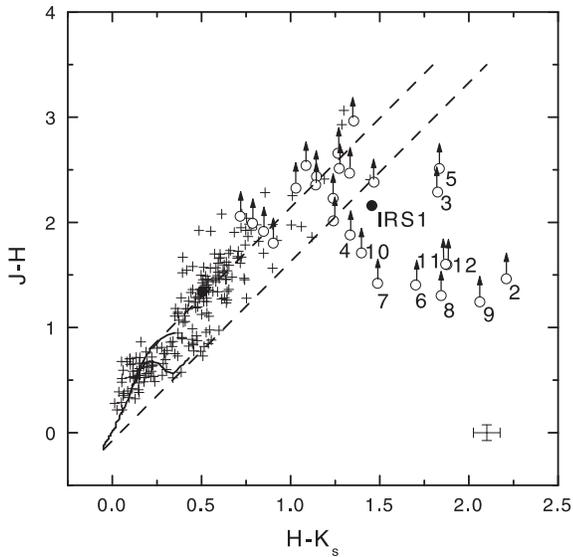,width=7.5cm,angle=0}
\end{center}
\caption[]{Color-color diagram of the sources extracted from the 2MASS data. The circles 
represent YSOs. The solid curves show the locus of the unreddened main sequence and giant 
stars, the two parallel dashed lines form the reddening band and confine the region in which 
stars with normal photosphere fall. A solid circle marks the positions of star IRS1 (=I\,20231). 
Sources not detected in $J$ band are shown as open circles. Arrows indicate the lower limits,
for which the $J$ band completeness limit of 17\,mag is adopted for $J$ magnitudes. YSO 
candidates are labeled with sequence number from 2 and 12 according to their offset from 
I\,20231. In the lower right corner the average photometric error cross is indicated.}
\label{20231-ccplot}
\end{figure}

Stars within 8$'$ diameter of I\,20231 are extracted from The Two Micron All Sky Survey 
(2MASS, see {\it http://www.ipac.caltech.edu/2mass/} for details) catalogue which contains
$J$, $H$ and $K_{\rm s}$ magnitudes of stars. Completeness limits in magnitude are close to 
17, 16 and 15\,mag in the $J$, $H$ and $K_{\rm s}$ bands, respectively. The $J-H/H-K_{\rm s}$ 
color-color diagram presented in Fig.\,\ref{20231-ccplot} includes stars detected at least in 
$H$ and $K_{\rm s}$ with magnitudes of 6.0$\leq$ {\it m($K_{\rm s}$)}$\leq$14.3 and photometric 
uncertainties less than 0.2\,mag in all detected bands. Sources satisfying the above criteria 
but not being detected in $J$ band are denoted with open circles. The completeness limit of
17\,mag is adopted as a lower limit of $J$ magnitudes for these sources. While the solid curve draws 
the locus of points corresponding to unreddened main sequence and giant branch stars (Koornneef 
\cite{Koornneef83}), the two parallel dashed lines form the reddening band and are bound to the 
region of stars with normal photospheres (see Lada \& Adams \cite{Lada92}). The sources lying 
to the right side of the reddening band are believed to be mostly YSOs (Lada \& Adams \cite{Lada92}; 
Lada et al. \cite{Lada93}). Careful scrutiny of the region reveals 12 YSO candidates (represented 
by circles labeled with sequence numbers in Fig.\ref{20231-ccplot}, see also Fig.\ref{20231+3440-bolo}). 
Since almost all these YSOs except for IRS1 (solid circle) were not detected in $J$, their 
positions in the color-color diagram are still uncertain. Deep photometric observations are 
therefore needed to confirm their YSO nature. Table\,\ref{2mass-tab} includes their near-infrared 
photometric data. From its location on the color-color diagram, we can estimate the extinction
of IRS1 to be $\sim$15, much lower than that derived from the peak emission of the 870$\mu$m 
dust continuum ($A_{\rm V}$\,$\sim$\,100). Although the $A_{\rm V}$ estimate using
dust emission is model dependent (see \S\,\ref{dust}), the difference in the extinction estimates
is most likely beyond the uncertainties induced by dust properties. This may suggest that 
the core is structured and/or that IRS1 is not at its center, in which case
SMM1 and IRS1 may not refer to the same source. This is further supported by the 
fact that IRS1 is located $\sim$8$''$ ($\sim$0.04\,pc) northwest 
of the 870$\mu$m peak (SMM1; see also \S\,\ref{driving}).

\begin{table}
\caption[]{\label{2mass-tab} 2MASS NIR properties of embedded sources}
\begin{tabular}{rclccc}
\hline
\hline
No.  & R.A.(2000) & DEC.(2000) & $J$ & $H$ & $K_{\rm s}$ \\
     &  (\,$h$\,\, $m$\,\, $s$) & (\,\,\,\,$^{\circ}$\,\,\, $'$\,\,\, $''$\,\,) & (mag) & (mag) & (mag) \\
\hline
\noalign{\smallskip}
IRS\,1 & 20 25 07.2 & +34 50 05 & 13.37 & 11.21 &  9.76 \\
  2 & 20 25 07.3 & +34 49 56 & -- & 15.54 & 13.33 \\
  3 & 20 25 04.8 & +34 50 14 & -- & 14.71 & 12.89 \\
  4 & 20 25 07.0 & +34 49 31 & -- & 15.12 & 13.79 \\
  5 & 20 25 04.3 & +34 50 08 & -- & 14.49 & 12.66 \\
  6 & 20 25 06.6 & +34 49 15 & -- & 15.60 & 13.90 \\
  7 & 20 25 07.4 & +34 51 23 & -- & 15.58 & 14.09 \\
  8 & 20 25 03.4 & +34 51 20 & -- & 15.70 & 13.85 \\
  9 & 20 24 58.3 & +34 49 39 & -- & 15.76 & 13.70 \\
  10 & 20 25 05.2 & +34 52 13 & -- & 15.29 & 13.89 \\
  11 & 20 25 05.6 & +34 47 36 & -- & 15.54 & 13.88 \\
  12 & 20 25 02.9 & +34 46 13 & -- & 15.40 & 13.53 \\
\hline
\end{tabular}
\end{table}

\section{Discussion}
\label{discussion}

\subsection{Physical properties of the two cores}
\label{physical}

Dense cores are the birthplaces of stars. Their study provides deep insights
into the initial conditions of cool dense gas clouds that determine important
stellar parameters like mass, metallicity, and initial rotation period. Jijina et al. 
(\cite{Jijina99}) compiled a catalog of 264 dense cores from \AMM\ observations, 
and found that the presence of cluster associations has a dramatic effect on core
gas and YSO properties. The line widths, kinetic temperatures and core sizes for 
cores in clusters are much larger than the corresponding physical parameters for 
cores in isolated star-forming environments.

The intrinsic line widths averaged over the \AMM\ cores are 1.79 and 1.20\,\kms\ 
for the northern and the southern source, respectively. These values are very close 
to the cluster associated median values of 1.20\,\kms\ from the Jijina samples 
(samples IV and V in Jijina \cite{Jijina99}) and larger than the median values for 
those samples without association (samples I, II and III in Jijina et al. 
\cite{Jijina99}). Moreover, turbulence in both cores is predominanantly non-thermal, 
since  for the molecule of mean mass (=1.36\,m$_{\rm H_{\rm 2}}$) thermal line widths $\Delta$$v_{\rm T}$ (0.63 and 0.49\,\kms\ for the 
northern and the southern core, respectively) determined from the kinetic temperature  
are small relative to the nonthermal line widths $\Delta$$v_{\rm NT}$ (1.77 and 
1.18\,\kms), obtained by subtracting in quadrature the thermal line width from 
the intrinsic line width averaged over the core. Further supported by substantial
core masses, the two \AMM\ cores are reminiscent of the `turbulent, massive dense 
cores' forming clusters or groups of stars in the Jijina et al. samples, although 
none of the two cores seems to have more than 30 embedded stars associated. This 
is supported by the distribution of the YSO candidates shown in Fig.\,\ref{20231+3440-bolo}, 
in which at least six (IRS1, 2 to 6) appear to be associated with the northern core 
and another two with the southern one (11 and 12).

Towards the northern core, there seems to be a temperature gradient from the core
center to the outermost layer. The detection of the (4,4) line of \AMM\ yields a
kinetic temperature of $\sim$30\,K at the position of I\,20231. This value is 
an average over the beam size of 40$''$ ($\sim$0.2\,pc) and is likely a lower 
limit for the region immediately surrounding I\,20231, since the (4,4) line was 
only detected towards the IRAS position and may not have been spatially 
resolved. The average kinetic temperature over the rest of the core is $\sim$\,15\,K. 
For regions where \AMM\ is detectable, the density is on the order of $10^{\rm 4}$\,\percc. 
At lower densities ($10^{\rm 2-3}$\,\percc), an excitation temperature of $\sim$\,11\,K
is traced by \CEIO, decreasing further to $\sim$\,7.5\,K towards the outermost layer
of the clouds (see \S\,\ref{ambient}).

\subsection{The driving sources of the outflows}
\label{driving}

The far-infrared luminosity of the IRAS source, calculated from the IRAS flux, 
is $L_{\rm FIR}$=196 $\solum$ (at a distance of 1\,kpc). This luminosity is 
higher than the minimum value set by the mechanical luminosity requirement. 
By adopting the $\dot{M}-L_{\rm bol}$ relationship of Shepherd \&\ Churchwell 
(\cite{Shepherd96}), the luminosity of the outflow source in I\,20231 is 
predicted to be greater than 114\,$\solum$. Within the error range, it suggests 
that the infrared source in question can provide sufficient mechanical power to 
drive the CO outflow.

Source IRS1, located at (3$''$, 0$''$), coincides well with the IRAS source 
I\,20231 with $K_{\rm s}$=9.76\,mag and $H-K_{\rm s}$=1.46\,mag. Immediately 
next to IRS1, a fainter ($K_{\rm s}$=13.33 mag) companion source 2 was found 
at (4$''$, --11$''$), appearing much redder with  $H-K_{\rm s}$=2.21 mag. 
Are they physically related? And which one is responsible for the outflow? 
Considering the clumpiness in outflow structure, are perhaps both stars 
driving the outflow? It is worth noting that the position difference between 
source 2 and SMM1 is 5$''$, which is marginally within the pointing uncertainty 
of the 870$\mu$m observations. The question then arises on how IRS1, source 2 
and SMM1 are physically related. Considering the relatively large beam (22$''$)
of the 870$\mu$m observations, higher resolution dust continuum and molecular 
line observations as well as some deeper NIR images are needed to answer 
these questions.

In the southern core, we did not find such YSO candidates close to SMM2.
Although there is a much weaker (14.13\,mag) $K_{\rm s}$ source located at 
(--3$''$,--180$''$), its nature is unknow since it was not detected
in $J$ and $H$ and we cannot place it on the color-color diagram. Most 
likely the driving source of the southern outflow is still deeply embedded 
in the cloud. It is also less luminous than its counterpart(s) of the northern 
outflow.

\subsection{Evolutionary status of the two cores}
\label{evolution}

Protostars evolve from Class\,0 to Class I, II, III. Class\,0 sources, 
the youngest protostars, are characterized by $M_{\rm env}/L_{\rm bol} > 
0.1$\,$\solmass/\solum$ ($M_{\rm env}$ is the envelope mass), no detectable 
emission at wavelengths shorter than $\sim$\,10\,$\mu$m, and the presence of 
outflow (Andr{\'e} et al. \cite{Andre93}, \cite{Andre00}). At present, 42 
Class\,0 sources have been identified (Andr{\'e} et al. \cite{Andre00}).

A reasonably good estimate of $M_{\rm env}$ can be provided by the strength 
of the optically thin 870\,$\mu$m continuum emission in the 22$''$ beam 
($\sim$\,0.1\,pc, the typical size of protostellar envelopes in the Taurus 
cloud). Following the method used in \S\,\ref{dust}, we can estimate $M_{\rm env}$ 
for SMM1 and SMM2 to be about 9 and 14\,\solmass, with their peak flux 
densities of 1.67 and 0.60\,Jy\,beam$^{-1}$, respectively. It is, however, 
worth noting that $\beta$ = 1.5 ($\kappa_{\rm 870\mu m}$ = 
0.02\,cm$^{\rm 2}$g$^{\rm -1}$ correspondingly), a value lower than the one used 
for mass calculations in \S\,\ref{dust}, is found to be more suitable to estimate the 
masses traced by the dust around protostars (e.g. Visser et al. \cite{Visser98}, 
\cite{Visser01}). Moreover, the averaged dust temperature within a 22$''$ 
beam could be higher than the ones used in \S\,\ref{dust}. Both effects lead to lower
masses. Considering additionally that I\,20231 may not be at the core center, 
we can safely conclude that $M_{\rm env}/L_{\rm bol}<0.04$\,\solmass/\solum\ 
for SMM1, typical of a Class\,I source. The detection of I\,20231 at 
12\,$\mu$m and 2\,$\mu$m and the moderate strength of its CO outflow also 
suggest SMM1 to be a $\sim$\,200\,\solum\ Class\,I YSO.

To better constrain the properties of SMM2, we first estimate the lower limit 
to $M_{\rm env}$. $\kappa_{\rm 870\mu m}$ = 0.03\,cm$^{2}$g$^{-1}$ is most likely an 
upper limit if we take the assumption from Hildebrand (\cite{Hildebrand83}), 
i.e. $\kappa_{\rm \lambda} = 0.1(250/ \lambda (\mu m))^{\rm \beta}$ cm$^{2}$g$^{-1}$, with 
$\beta$ = 1, which is the lower limit in star forming regions at the millimeter and 
submillimeter wavelengths. A dust 
temperature of 30\,K is most likely an upper limit to SMM2. These parameters
yield a lower limit to the envelope mass of 1\,\solmass. Meanwhile, since 
there is no detectable IRAS source associated with SMM2, the sensitivity 
of the IRAS instruments (0.5\,Jy at 12, 25, and 60\,$\mu$m, 1.5\,Jy at 
100\,$\mu$m) allow us to estimate an upper limit of $L_{\rm bol}\sim$\,6\,\solum\ 
for the luminosity of the possible source(s) embedded in the southern core. 
The $M_{\rm env}/L_{\rm bol}$ ratio for SMM2 is therefore comfortably greater than 
0.1\,\solmass/\solum, typical of a Class\,0 source. Combined with the fact 
that SMM2 also harbors a bipolar outflow, the southern core is more likely
to contain a Class\,0 object with 1\,\solmass\,$\la$ M$_{\rm env}$$\la$\,14\,\solmass.

Since outflows are found in both cores, we can also compare the evolutionary 
phases of the two cores by placing these two sources on the outflow efficiency 
versus $M_{\rm env}/L_{\rm bol}^{0.6}$ diagram (see fig. 7 of Bontemps et al. 
\cite{Bontemps96}). The outflow efficiency is the outflow force divided by the 
radiative momentum flux ($L_{\rm bol}/c$) and is  $\sim$\,80 (or $\sim$\,380 
also accounting for the low velocity flow traced by \HCOP) for the northern 
outflow and $>$220 for the southern one. Accounting for the different 
$M_{\rm env}/L_{\rm bol}^{0.6}$ values of the two cores, the northern one lies 
in the transit region between Class\,0 and I, while the southern one lies 
distinctly in the region occupied by Class\,0 sources. This, in agreement
with the fact that the dynamical time of the northern flow is longer than that 
of the southern flow (see \S\,\ref{outflow}), further supports the protostellar 
status of SMM1 (I\,20231) and SMM2. 

We thus conclude that the southern core is in an earlier evolutionary stage 
than the northern core. One may ask why there is no water maser detection 
in the southern core, since Class\,0 sources are  believed to be favorable 
sites for such masers (Furuya et al. \cite{Furuya01}). An answer could be
that the maser is too weak to be detected at an rms level of 0.1\,Jy, since 
according to the $L_{\rm H_{\rm2}O}-L_{\rm bol}$ correlation found by Furuya 
et al. (\cite{Furuya01}), we expect for a protostar with 
$L_{\rm bol}<6\,\solum$ a luminosity of $L_{\rm H_{\rm2}O}\la 1.7\times 10^{-11}\,\solum$.
This is about two orders of magnitude below the estimated upper limit (see \S\,\ref{water}).  
On the other hand, the non-detection of a \WAT\ maser in the southern core 
is not inconsistent with the recent statistical study by Zhang et al. 
(\cite{Zhang01}) who found that the outflow may develop before the appearance 
of a water maser although their study is based on a sample of massive star 
forming regions. It is therefore also possible that the southern core has 
not yet reached an evolutionary stage to form a luminoous \WAT\ maser, while 
an outflow was already activated.

The above disscussion is based on the assumption that SMM1 and IRS1 refer to
the same source as I\,20231, but this is not necessarily the case. As we 
have mentioned in \S\,\ref{2mass} and \S\,\ref{driving}, the differences in extinction ($A_{\rm v}$ $\sim$\,100 
versus $\sim$\,15) and in position between SMM1 and IRS1 ($\sim$\,8$''$ or 0.04\,pc)
raise the possibility that the two sources are distinct objects of the northern
``cluster". In such a case, L$_{\rm bol}$ of  SMM1 would be significantly 
lower than the value estimated from FIR flux 
densities of I\,20231, and as a result SMM1 might still be a Class\,0 object 
and driving source of the northern outflow. High angular
resolution observations would be required to draw definite conclusions.

\subsection{Velocity gradients and the asymmetric line profiles}
\label{infall}

\begin{figure}
\vspace{0.3cm}
\psfig{figure=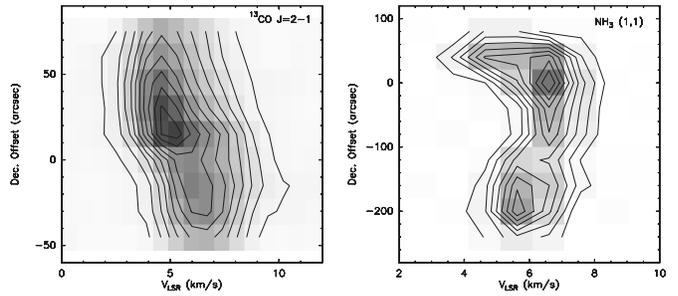,width=8.5cm,angle=0}
\caption[]{Position--velocity maps of \THCO\ $J$ = 2--1 (left) and \AMM\,(1,1)  
(right) emission
along Dec. for R.A. offset = 0. The contours are 10\% to 90\% by 10\% of the 
peak emission (9.2\,K and 4.1\,K for \THCO\ $J$ = 2--1 and \AMM\,(1,1), 
respectively).}
\label{20231-pv}
\end{figure}

There exists a velocity gradient from north to south in all molecular spectra, 
indicating an internal motion. In the case of the \AMM\ lines, the velocity
increases monotonously from 5.2\,\kms\ at (0$''$,80$''$) to 6.8\,\kms\ at 
(0$''$,--80$''$), where there seems to be a turning point. Further south the 
velocity shifts back to and stays almost constant at 6.0\,\kms. Such a gradient 
is also found in \THCO\ $J$ = 2--1, where the velocity shifts from 5\,\kms\ at 
(0$''$,75$''$) to 6.3\,\kms\ at (0$''$,-45$''$). The CS $J$ = 5--4 line profile
at the central position is quite similar to that of the \THCO\ line, with a peak 
slightly skewed to low velocity. Again, the velocity increase from 4.5 \kms\ at 
(0$''$,75$''$) to 6.5 \kms\ at (0$''$,-30$''$), and drops to 6.0 \kms\ further 
south just like ammonia. This  indicates that a boundary exists near (0$''$,-100$''$), 
separating the northern cloud (with a `systemic' velocity $\sim$ 6.3\,\kms) from 
the southern cloud (with $\sim$ 6.0\,\kms). The gradients across the northern 
core are 2 to 4\,\kmspc, slightly larger than the typical rotation induced gradients 
of 1\,\kmspc\ for molecular cloud cores (Goodman et al. \cite{Goodman93}), but 
within the range found by Jijina et al. (\cite{Jijina99}) in their sample.  As 
examples, we present in Fig.\,\ref{20231-pv} the position-velocity maps of \THCO\ 
$J$ = 2--1 (left) and \AMM\,(1,1) (right). Morphologically, the diagrams
are consistent with a large cloud that shows overall large scale rotation.
However, such an effect could also be caused by two independent cores with slightly 
different velocities.

The \HCOP\ $J$ = 3--2 profiles are of particular interest, owing to the presence 
of strong blue asymmetry (see \S\,\ref{hco+}), indicating possible infall motion (Leung 
\& Brown \cite{Leung77}; Zhou \cite{Zhou92}; Gregersen et al. \cite{Gregersen97}). 
The spatial extent ($\sim$ 0.2\,pc) of such an asymmetry is not inconsistent 
with the finding of Myers et al. \cite{Myers00}, who detected infall motion over 
a linear scale of $\sim$\,0.1\,pc in a large number of starless cores and Class 0/I 
protostars. However, it is hard to study the detailed infall structure with current 
spatial resolution. Contamination by rotation and outflows make a clearcut
interpretation even more difficult. 
Further high spatial resolution observations are imperative to verify such a 
possible infall.

\section{Conclusions}
\label{conclusion}

We have presented detailed molecular line and sub-mm dust observations
of the molecular cloud \object{Lynds\,870}, a previously poorly studied region. 
The main results are:

1. Substantial molecular emission has been detected from \object{Lynds\,870}.
Two main cores, separated by 3$'$, are found both in dust (SMM1 and SMM2) 
and molecular line emission with the northern core (SMM1) being particularly 
rich in molecular lines. The continuum and spectroscopic observations indicate that 
the northern core, centered at \object{IRAS\,20231+3440}, is warmer and denser 
than  the southern core (SMM2). A total mass of $\sim$70--110\,\solmass\ is 
estimated from the molecular line and the submillimeter dust emission. 

2. Outflow activities are detected in both the northern and the southern
core, indicative of on-going star formation in both regions. While 
\object{IRAS\,20231+3440} (IRS1 in $K_{\rm s}$) is found to be a Class\,I 
YSO candidate and most likely the exciting star of the northern outflow, 
a more deeply embedded companion (source 2 in $K_{\rm s}$) found next to 
\object{IRAS\,20231+3440} may also be a possible driving engine. The driving 
source of the southern outflow, still deeply embedded in the cloud,
is most likely a Class\,0 YSO candidate. The southern core (SMM2) is therefore 
still in an earlier evolutionary stage than the northern core (SMM1).

3. A significant amount of low velocity outflowing gas is hidden in the relatively
broad systemic CO profiles tracing the quiescent gas. The low velocity outflow is, 
however, revealed by the \HCOP\ emission. The total outflowing gas mass of the 
northern core is derived to be 8\,\solmass. 

4. Possible infall motions in the northern dense core are indicated by 
self-absorbed \HCOP\,$J$ = 3--2 line profiles with a strong blue asymmetry; 
its spatial extent is $\sim$\,0.2\,pc.

5. In a diameter of 8$'$ around \object{IRAS\,20231+3440}, there are in total 
12 $K_{\rm s}$ sources with a strong IR excess, typical of YSOs still embedded 
in their parental dust/molecular cloud. Combined with the outflows and \WAT\ 
masers found in the region, we conclude that the cloud is presently actively 
forming a group or a small cluster of stars. 

\acknowledgements{We appreciate the assistance of the HHT staff and Effelsberg-100m operators
during the observations. We are grateful to Dr. D. Muders 
for his help in the reduction of HHT 870\,$\mu$m continuum data. 
We also benefited from useful discussions with Drs. Q. Zeng, G. Sandell, and Q. Zhang.
We thank the referee whose helpful comments and suggestions greatly helped to 
improved the paper.
This publication makes use of data products from the Two Micron All Sky Survey, 
which is a joint project of the University of Massachusetts and the Infrared 
Processing and Analysis Center, funded by the National Aeronautics and Space 
Administration and the National Science Foundation. R.Q.M. acknowledges support 
by the exchange program between the Chinese Academy of Sciences and the 
Max-Planck-Gesellschaft. 
This work was supported in part by grants 19625307 from NSFC \& G19990754 from CMST.}

\end{document}